\def\year{2019}\relax
\documentclass[letterpaper]{article} %DO NOT CHANGE THIS
\usepackage{aaai19}  %Required
\usepackage{times}  %Required
\usepackage{helvet}  %Required
\usepackage{courier}  %Required
\usepackage{url}  %Required
\usepackage{graphicx}  %Required
\usepackage{algorithm}
\usepackage{algorithmicx}
\usepackage{algpseudocode}
\usepackage{amsmath}
\usepackage{float}
\usepackage{lipsum}

\begin{document}
\title{RADMPC: A Fast Decentralized Approach for Chance-Constrained Multi-Vehicle Path-Planning}
\author{Aaron Huang, Benjamin J. Ayton, Brian C. Williams\\
Computer Science and Artificial Intelligence Laboratory\\
Massachusetts Institute of Technology\\
32 Vasser Street\\
Cambridge, MA 02139\\
}
\maketitle
\begin{abstract}
Robust multi-vehicle path-planning is important for ensuring the safety of multi-vehicle systems in applications like transportation, search and rescue, and robotic exploration. Chance-constrained methods like Iterative Risk Allocation (IRA)\cite{IRA} have been developed for situations where environmental disturbances are unbounded. However, chance-constrained methods for the multi-vehicle case generally use centralized strategies where the vehicle set is planned with couplings between all vehicle pairs. This approach is intractable as fleet size increases because computation time is exponential with respect to the number of vehicles being planned over due to a polynomial increase in coupling constraints between vehicle pairs. We present a faster approach for chance-constrained multi-vehicle path-planning that relies upon a decentralized path-planning method called Risk-Aware Decentralized Model Predictive Control (RADMPC) to rapidly approximate a centralized IRA approach. The RADMPC approximation is evaluated for vehicle interactions to determine the vehicle sets that should be planned in a coupled manner. Applying IRA to the smaller vehicle sets determined from the RADMPC approximation rapidly plans safe paths for the entire fleet. A Monte Carlo simulation analysis demonstrates the correctness of our approach and a significant improvement in computation time compared to a centralized IRA approach.
\end{abstract}

\section{Introduction}
Recent interest in the utilization of autonomous vehicle fleets has skyrocketed for applications like urban transportation and undersea exploration. There is an imminent need for architectures that support safe coordination of multiple vehicles in a practical and optimal manner. A critical element of multi-vehicle coordination is safe multi-vehicle path-planning in an environment where collisions are possible. Paths for autonomous vehicle fleets should be optimal by some measure (e.g. actuation cost, solution execution time) and must have a reasonable computation time to be practical.

Prior methods for fast multi-vehicle path-planning are primarily fixated on situations where environmental disturbances are assumed to be bounded. Robust Model Predictive Control (RMPC) techniques have been extensively studied for multi-vehicle path-planning in this case. However, the assumption of bounded environmental disturbances is invalid in many practical applications. Chance-constrained techniques are commonly used for path-planning in the alternate case where environmental disturbances are assumed to be unbounded. Current chance-constrained techniques for the multi-vehicle case generally use a \emph{centralized} strategy where the couplings are present for all vehicle pairs in the entire vehicle set. Centralized strategies are intractable as fleet size increases because computation time is typically exponential as more vehicles are considered. This is due to a combinatorial increase in binary variables that are required for vehicle and obstacle avoidance constraints.

We present a faster approach to chance-constrained multi-vehicle path-planning that relies upon a novel method called Risk-Aware Decentralized Model Predictive Control (RADMPC). RADMPC extends from Decentralized Model Predictive Control (DMPC) techniques by using a single iteration of IRA as the primary optimization method. RADMPC plays a key role in reducing the complexity of the multi-vehicle path-planning problem by rapidly approximating a centralized application of IRA. The approximation is examined for vehicle interactions to determine smaller sets of vehicles that need to be planned in a coupled manner. Individually applying IRA to the smaller vehicle sets plans safe paths for the vehicle fleet much more quickly than a centralized IRA approach.

\section{Literature Review}
There is a considerable body of work in path-planning for multiple autonomous vehicles. This is a hard problem because of two reasons: the presence of environmental uncertainty and the nonconvexity of the optimization problem. Mixed-Integer Linear Programming \cite{MILP} and Disjunctive Linear Programming \cite{DLP} strategies were used in initial techniques to handle the nonconvex optimization problems. However, these approaches did not account for sources of uncertainty in the problem.

Many methods that handle uncertainty usually make the assumption of bounded uncertainties and environmental disturbances, allowing for the design of robust trajectories that are resistant to constraint failure against the worst case disturbances. Robust Model Predictive Control (RMPC) is an extension of Model Prediction Control (MPC) that is commonly used for path-planning under this assumption. Trajectories generated with RMPC techniques have been shown to be safe from obstacle collision at up to 3$\sigma$ confidence at each timestep \cite{RMPC-RRT,RMPC,RMPC-SURVEY}. However, RMPC techniques are still intractable for multi-vehicle problems with larger groups of vehicles.
	
Decentralized Model Predictive Control (DMPC) \cite{DMPC} algorithms employ the strategy of decomposing the full multi-vehicle trajectory optimization problem into decentralized subproblems to reduce computational intensity. Each subproblem optimizes the trajectory of a single vehicle using an RMPC strategy. To account for inter-vehicle coupling constraints, each subproblem must be solved while considering all other vehicle trajectories. DMPC techniques are computationally inexpensive compared to RMPC techniques for multi-vehicle problems. 

Many real-life situations involve uncertainties that cannot be bounded, making it impossible to guarantee constraint satisfaction with zero probability of failure using most RMPC and DMPC approaches. In lieu of guaranteed constraint satisfaction, a different approach to path-planning is to place \emph{chance constraints} to limit the \emph{probability} of violating constraints. There are two kinds of chance constraints: \emph{individual} chance constraints limiting the probability of failure of a single constraint at a single timestep, and \emph{joint} chance constraints limiting the probability of failure of any constraint in a problem \cite{PROBABILISTIC-MPC}. 

Solving a multi-vehicle path-planning problem with a joint chance constraint is hard because it requires the computationally intensive calculation of the probabilities of non-independent events. \citeauthor{BLACKMORE-ONO-WILLIAMS} demonstrate an elegant method in which Boole's inequality is used to decompose the joint chance constraint into individual chance constraints. The new RMPC problem with individual constraints can be more easily solved by constraint tightening. However, this chance-constrained method is conservative since it assigns a uniform value to the risk bound for every individual chance constraint. 

Iterative Risk Allocation (IRA) \cite{IRA} is a two-stage optimization method that uses the concept of \emph{risk reallocation} to reduce the suboptimality of chance-constrained approaches. By iteratively reallocating risk from \emph{inactive} chance constraints to \emph{active} chance constraints, the allocation of risk that optimizes the objective function can be found. Unfortunately, the centralized application of IRA still has an exponential solution time and is impractical for problems with many vehicles.

A strategy for reducing the complexity of centralized approaches is to decouple unnecessary coupling constraints between vehicle pairs using heuristics. \citeauthor{KEVICZKY-ETAL} uses a distance-based heuristic for multi-vehicle path-planning to maintain a communication topology graph that is updated over time. Undirected edges between any two vehicles represent a coupling constraint indicating that either vehicle must account for the other vehicle's actions when planning. However, the distance-based heuristic does not make full use of the vehicle information available when determining coupling constraints.

\section{Problem Statement}
\subsection{Notation}
\noindent The following notation is used throughout the paper
\begin{align*}
    \boldsymbol{x}_i^k&:& &\text{State vector for vehicle } i \text{ at time } k\\
    \boldsymbol{u}_i^k&:& &\text{Control input for vehicle } i \text{ at time } k\\
    \boldsymbol{w}_i^k&:& &\text{Disturbance for vehicle } i \text{ at time } k\\
    \bar{\boldsymbol{x}}_i^k&:=&& E[\boldsymbol{x}_i^k]: \text{Nominal state for vehicle } i \text{ at time } k\\
    \delta_j^k&:& &\text{Risk bound for chance constraint } j \text{ at time } k\\
    \Delta&:& &\text{Risk bound for the joint chance constraint}
\end{align*}

\begin{align*}
\boldsymbol{X} := \begin{bmatrix}
\boldsymbol{x^0_0} \\
\vdots \\
\boldsymbol{x_N^T}
\end{bmatrix}
&&&
\boldsymbol{\bar{X}} := \begin{bmatrix}
\boldsymbol{\bar{x}^0_0} \\
\vdots \\
\boldsymbol{\bar{x}_N^T}
\end{bmatrix}
&&&
\boldsymbol{U} := \begin{bmatrix}
\boldsymbol{u^0_0} \\
\vdots \\
\boldsymbol{u_N^T}
\end{bmatrix} \\\\
    i \in [0 \hdots N] &&& j \in [0 \hdots L] &&& k \in [0 \hdots T]
\end{align*}

\noindent $N$ denotes the number of vehicles to plan over. $T$ denotes the total number of timesteps in the problem. $L$ refers to the number of chance constraints present.

\subsection{RMPC with a joint chance constraint}
The chance-constrained multi-vehicle path-planning problem is formulated as follows:
\begin{align}
& \underset{\boldsymbol{U}}{\text{min}} & & E[J(\boldsymbol{X},\boldsymbol{U})] \\
& \text{s.t.} & & \boldsymbol{x}_i^{k+1} = A\boldsymbol{x}_i^k + B\boldsymbol{u}_i^k + \boldsymbol{w}_i^k \\
& & & \boldsymbol{u}_{i,\text{min}} \leq \boldsymbol{u}_i^k \leq \boldsymbol{u}_{i,\text{max}} \\
& & & \boldsymbol{w}_i^k \sim \mathcal{N}(0, \Sigma_{w_i^0}) \\
& & & \boldsymbol{x}_i^0 \sim \mathcal{N}(\bar{\boldsymbol{x}}_i^0, \Sigma_{x_i^0}) \\
& & & \text{Pr}\Bigg[\overset{N}{\underset{i=0}{\bigwedge}}\overset{L}{\underset{j=0}{\bigwedge}}\overset{T}{\underset{k=0}{\bigwedge}} \boldsymbol{h}_j^{kT} \boldsymbol{x}_i^k \leq g_j^k\Bigg] \geq 1 - \Delta
\end{align}

We model vehicles as discrete-time linear time invariant (LTI) systems operating in the presence of \emph{unbounded} environmental disturbances. In Equation 1, we wish to determine the control sequence $\boldsymbol{U}$ that generates a state sequence $\boldsymbol{X}$ that minimizes the expected value of the objective function $\boldsymbol{J}$ while obeying Equations 2 - 6. Equation 2 defines state evolution for vehicles from time $k$ to time $k+1$ where matrices $A$ and $B$ represent linear vehicle dynamics and control effects. The combination of Equation 2 and Equation 4 explicitly describe the effects of an unbounded Gaussian disturbance $\boldsymbol{w_i^k}$, forcing vehicle states to be unbounded. Equation 6 defines the total risk bound $\Delta$ as the upper bound on the probability that any individual chance constraint fails. 

\section{Preliminaries}
We will briefly describe IRA and the decomposition of the joint chance constraint into individual chance constraints to provide better context for our approach.

\subsection{RMPC with individual chance constraints}
We cannot easily solve RMPC with a joint chance constraint because Equation 6 involves the integration of a multivariate Gaussian distribution. However, this problem can be made more tractable by decomposing the joint chance constraint into individual chance constraints using Boole's inequality ($\text{Pr}[A\cup B] \leq \text{Pr}[A] + \text{Pr}[B]$).

\begin{align} 
\nonumber & \underset{\boldsymbol{U}}{\text{min}} & & E[J(\boldsymbol{X},\boldsymbol{U})] \\ 
\nonumber & \text{s.t.} & & \boldsymbol{x}_i^{k+1} = A\boldsymbol{x}_i^k + B\boldsymbol{u}_i^k + \boldsymbol{w}_i^k \\ 
\nonumber & & & \boldsymbol{u}_{i,\text{min}} \leq \boldsymbol{u}_i^k \leq \boldsymbol{u}_{i,\text{max}} \\ 
\nonumber & & & \boldsymbol{w}_i^k \sim \mathcal{N}(0, \Sigma_{w_i^0}) \\ 
\nonumber & & & \boldsymbol{x}_i^0 \sim \mathcal{N}(\bar{\boldsymbol{x}}_i^0, \Sigma_{x_i^0}) \\ 
& \forall i,j,k & & \text{Pr}\Big[\boldsymbol{h}_j^{kT}\boldsymbol{x}_i^k \leq g_j^k\Big] \geq 1 - \delta_j^k \\
& & & \overset{L}{\underset{j = 0}{\sum}}\overset{T}{\underset{k = 0}{\sum}} \delta_j^k \leq \Delta
\end{align}

Decomposing the joint chance constraint directly reduces Equation 6 to Equation 7, with Equation 8 constraining the sum of all $\delta_j^k$ to be no greater than $\Delta$. \citeauthor{BLACKMORE-ONO-WILLIAMS} demonstrate a method to solve this problem that transforms the stochastic problem into a deterministic problem. However, their method produces a suboptimal solution since it fixes all $\delta_j^k$ using a uniform allocation of $\Delta$. In other words, there may be an allocation of $\delta_j^k$ that produces a more optimal solution.

\subsection{Iterative Risk Allocation}
IRA is a method that is designed to achieve greater solution optimality for RMPC problems with joint chance constraints. IRA is a two-stage optimization method that uses \emph{risk reallocation} to determine the best risk allocation of $\delta_j^k$ that optimizes the objective function. The innovation of risk reallocation is in moving risk from inactive constraints to active constraints to monotonically decrease overall cost. 

To determine if a constraint is active or inactive, we must compare an individual chance constraint's risk bound to the probability of constraint satisfaction. This requires the probability of constraint satisfaction to be reformulated using deterministic variables. Recall that an individual chance constraint is defined as follows:

\begin{align*}
    P(\boldsymbol{h}_j^{kT}\boldsymbol{x}_i^k \leq g_j^k) &\geq 1 - \delta_j^k
\end{align*}

The lower stage optimization in IRA computes a solution for $\boldsymbol{x}_i^k$ using the fixed $\delta_j^k$ for each constraint. Then, we can define a constraint for the acceptable values of $\delta_j^k$ by writing the probability of constraint satisfaction in terms of a deterministic cumulative distribution function of $\boldsymbol{x}_i^k$.

\begin{align*}
    P(\boldsymbol{h}_j^{kT}\boldsymbol{x}_i^k \leq g_j^k) &= \text{cdf}(g_j^k - \boldsymbol{h}_j^{kT}\boldsymbol{x}_i^k) \\
    \implies \delta_j^k &\geq 1 - \text{cdf}(g_j^k - \boldsymbol{h}_j^{kT}\boldsymbol{x}_i^k)
\end{align*}

This result defines the minimum value for $\delta_j^k$ that is required to satisfy the chance constraint given the current solution for $\boldsymbol{x}_i^k$. Active and inactive constraints are determined by comparing the fixed $\delta_j^k$ to the newly defined $\delta_{j,min}^k$ using a tolerance $\eta$.

\begin{align*}
    \delta_{j,min}^k = 1 - \text{cdf}(g_j^k - \boldsymbol{h}_j^{kT}\boldsymbol{x}_i^k)\\
    \text{Active:}\:|\delta_j^k - \delta_{j,min}^k| \leq \eta\\
    \text{Inactive:}\: |\delta_j^k - \delta_{j,min}^k| > \eta
\end{align*}

The risk reallocation upper stage reallocates risk from active to inactive constraints while respecting the minimum risk bound. This defines a new risk allocation to be used in the next iteration of the lower optimization stage. This two-stage procedure is iteratively run for a fixed number of iterations or until the value of the objective function converges. Although this method produces excellent results, a centralized application of IRA to large multi-vehicle problems is computationally intensive because the number of avoidance constraints increases combinatorially.

\subsection{Avoidance Constraints}
Vehicle and obstacle avoidance constraints are defined by a disjunction of individual chance constraints each represented as a linear inequality. For a set of individual chance constraints that comprise an object $O$ to be avoided, the following disjunction of inequalities define a safe zone.

\begin{align*}
    \overset{O}{{\underset{j}{\bigvee}}}\;\boldsymbol{h}_j^{kT} \boldsymbol{x}_i^k \leq g_j^k
\end{align*}

However, using a disjunction of linear inequalities for obstacle avoidance is insufficient for path-planning because a vehicle cannot simultaneously be on every side of an object at once. Instead, we use a conjunction of the same linear inequalities and introduce binary variables $b_k^j$ that are multiplied with an arbitrarily large number $M$ to turn off boundary inequalities as needed. To ensure that at least one constraint is turned on, the sum of all binary variables must be less than the number of individual chance constraints that comprise $O$.

\begin{align*}
    \overset{O}{\underset{j}{\bigwedge}}\,\boldsymbol{h}_j^{kT} \boldsymbol{x}_i^k &\leq g_j^k + Mb_j^k \\
    \overset{O}{\underset{j}{\sum}} b_j^k &\leq card(O) - 1
\end{align*}

This formulation induces a polynomial increase in binary variables as more objects are considered. Vehicle avoidance constraints are similarly encoded. However, computation time is more adversely affected by additional vehicles than additional obstacles. This follows because avoidance constraints must exist between every vehicle pair, which grows combinatorially as more vehicles are added. A problem with $N$ vehicles must include ${N\choose2}$ vehicle avoidance constraints over $T$ timesteps. This implies that binary variables for vehicles increase with $O(N^2T)$ complexity and force an $O(e^{N^2T})$ solution time. On the other hand, additional obstacles increase binary variables with $O(NT)$ complexity, adding only $O(e^{NT})$ complexity. This distinction is important as we will later use a special form of obstacle called a \emph{temporal obstacle} to represent vehicles, reducing RADMPC complexity.

\section{Technical Approach}
We propose a three-step approach to produce the result of a centralized application of IRA to a large vehicle set with much less computational overhead. We present RADMPC - a fast, risk-aware path-planner used to approximate centralized IRA. The RADMPC approximation is evaluated to identify vehicle pairs with high probability of collision and decompose the full set of vehicles into smaller vehicle subsets that have only \emph{relevant} vehicle coupling constraints. Finally, we apply IRA to the smaller subsets to produce paths for all vehicles far more quickly than centralized IRA. Figure 1 depicts the flow of our approach. Figure 2 depicts example plots of different stages of our approach. The runtime of the centralized IRA solution exceeds the combined runtime of the RADMPC approximation and the runtime of applying IRA to the two vehicle subsets.

\begin{figure}[h]
\begin{center}
\includegraphics[width=1.0\columnwidth]{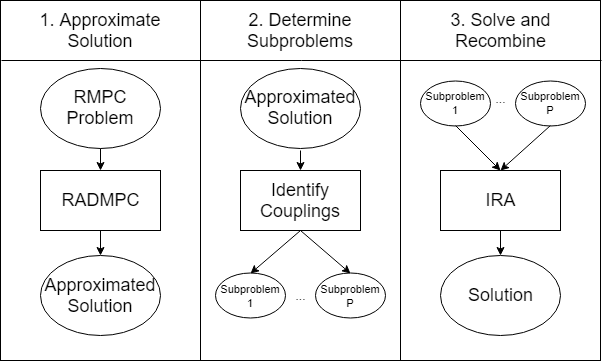}
\caption{The three stages of our approach}
\label{radmpc_flow}
\end{center}
\end{figure}

\begin{algorithm}
\caption{Multi-Vehicle}
\begin{algorithmic}[1]
\Procedure{Multi-Vehicle}{$V,\,N,\,T,\,\Delta,\,\Psi$}
\State{$\boldsymbol{X}_{appr} \gets$ RADMPC($V,\,N,\,T,\,\Delta$)}
\State{$V^* \gets$ findCouplings($\boldsymbol{X}_{appr},\,\Psi$)}
\For{each coupled set $v^*$ in $V*$ }
\State{$\Delta^* \gets \Delta \,\cdot$ card$(v^*)/ N$}
\State{IRA($v^*,\,N,\,T,\,\Delta^*$)}
\EndFor
\EndProcedure
\end{algorithmic}
\end{algorithm}

\subsection{RADMPC}

We have developed a novel path-planning algorithm called Risk Aware Decentralized Model Predictive Control (RADMPC) to be used as the fast path-planner in our approach. To be useful, it must be able to closely approximate centralized IRA. To be practical, it cannot be computationally expensive lest we fail to handle the core issue of solution time complexity for multi-vehicle path-planning. RADMPC is an extension of Decentralized Model Predictive Control (DMPC) as proposed by \citeauthor{DMPC}. However, our approach differs from DMPC by assuming unbounded environmental disturbances as opposed to bounded disturbances and using \emph{temporal obstacles} to communicate vehicle plans across subproblems.

RADMPC uses a decentralized path-planning strategy to quickly plan paths. Given $N$ vehicles, RADMPC decomposes the full problem into $N$ subproblems that each optimize the trajectory of a single vehicle. At time $k$, the subproblems are solved in a randomized order. The optimization method used to solve each subproblem simply consists of a single iteration of IRA with a uniform initial allocation over a \emph{risk pool}. After a subproblem is solved, the first control input of the solution is executed and a temporal obstacle is created to bound the vehicle's subproblem solution. RADMPC is recursively executed in this way until all vehicles are in their respective goal.

\begin{algorithm}
\caption{RADMPC}\label{euclid}
\begin{algorithmic}[1]
\Procedure{RADMPC}{$V,\,N,\,T,\,\Delta$}
\State $O \gets$ initTemporalObstacles($V$)
\State $\Delta^\# \gets \Delta,\,N^* \gets N$
\For{$k$ in $[0 \hdots T - 1]$}
\For{$i$ in random ordering of $[0 \hdots N - 1]$}
\State{$\boldsymbol{X}_{i}^*,\boldsymbol{\delta}_{i}^* \gets $fastIRA$(V_i,\,T - k,\,\Delta^\#/N^*)$}
\State{$\boldsymbol{x}_{i,appr}^{k+1} \gets \boldsymbol{x^*}_{i}^1$}
\If{all vehicles in goal}
\State \textbf{return} $\boldsymbol{X}_{appr}$
\ElsIf{$V_i$ in goal}
\State{$N^* \gets N^* - 1$}
\EndIf
\State{$O_i \gets $makeTemporalObstacle$(\boldsymbol{X}_{i}^*)$}
\State{$\Delta^\# \gets \Delta^\# -$ sum$(\boldsymbol{\delta^*}_i^1)$}
\EndFor
\EndFor
\EndProcedure{}
\end{algorithmic}
\end{algorithm}

\begin{figure*}[ht!]
\minipage{0.33\textwidth}%
\includegraphics[width=\linewidth]{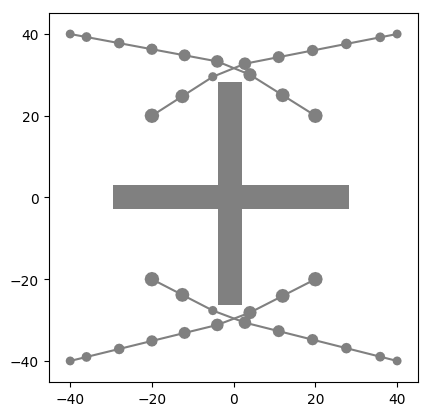}
\endminipage\hfill
\minipage{0.33\textwidth}
\includegraphics[width=\linewidth]{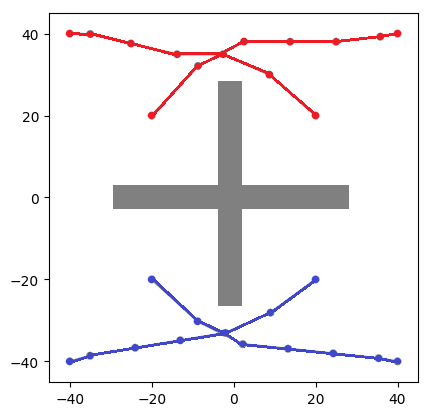}
\endminipage\hfill
\minipage{0.33 \textwidth}
\includegraphics[width=\linewidth]{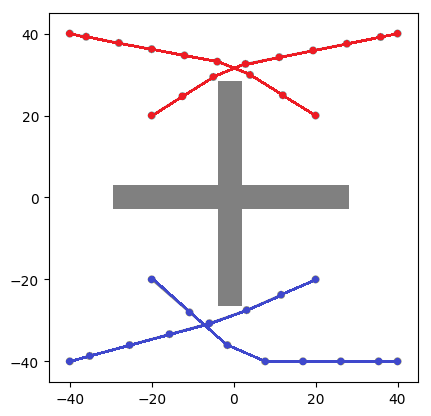}
\endminipage

\caption{Left to right: centralized IRA solution, RADMPC approximation, RADMPC solution. Runtime of the RADMPC approximation and IRA on coupled vehicle sets is generally much faster than runtime of centralized IRA.}
\label{vehicle_obstacles}
\end{figure*}

\subsubsection{Risk Pooling}
A \emph{risk pool} $\Delta^\#$ is used to track the total risk remaining while executing RADMPC. $\Delta^\#$ is initialized to $\Delta$. Every subproblem is given a uniform risk allocation over $\Delta^\#/N^*$ where $N^*$ is the number of vehicles that are not in goal. After a subproblem is solved, $\Delta^\#$ is updated as follows:

\begin{align*}
    \Delta^\# \leftarrow \Delta^\# - \overset{L}{\underset{j=0}{\sum}} \delta_0^j
\end{align*}

This allows RADMPC to be risk-aware by subtracting the risk that is used at the \emph{executed} timestep from the risk pool. The risk pool is then divided among the remaining vehicles for the next subproblem.

\subsubsection{Temporal Obstacles}
We introduce a \emph{temporal obstacle} representation for a vehicle's plan according to its most recently solved subproblem (hereafter referred to as \emph{temporal vehicle obstacles}. Temporal obstacles are a simple extension of static obstacles where the linear inequalities describing the obstacle boundaries change with respect to time.  

Temporal vehicle obstacles are created after a subproblem is solved for a vehicle $i$ using the subproblem solution $\boldsymbol{X}_i^*$ A temporal vehicle obstacle bounds the $3\sigma$ confidence region of $\boldsymbol{x^*}_i^k$ for all timesteps $k \in [0 \hdots T]$ for vehicle $i$. We define distance $d$ as the radial distance from $\bar{x^*}_i^k$ bounding the $3\sigma$ confidence region. As truly circular obstacles are difficult to represent, $d$ is instead used to compute square coordinates around $\boldsymbol{x^*}_i^k$. Square obstacles are used to represent vehicles in goal and vehicles at time $k = 0$. At other timesteps, the obstacle is generated by wrapping a convex hull around the square coordinates generated at $\boldsymbol{x^*}_i^k$ and $\boldsymbol{x^*}_i^{k+1}$.

Using a temporal obstacle representation for vehicles offers a light, yet powerful representation when other vehicles need to consider their intent while planning. Representing vehicles as temporal obstacles has $O(e^{NT})$ computational complexity as opposed to $O(e^{N^2T})$ complexity otherwise. This formulation also offers an intuitive way of evaluating interactions between vehicles planned using RADMPC.

\subsection{Identifying Couplings}
Couplings are determined by evaluating the probability of collision between vehicles and temporal vehicle obstacles. For each boundary in a temporal vehicle obstacle that has an activated linear constraint (i.e. the associated binary variable is  0), the collision probability is computed as follows \cite{BLACKMORE-ONO-WILLIAMS}:

\begin{align*}
    \text{Pr}\Big[\boldsymbol{h}_j^{kT}\boldsymbol{x}_i^k > g_j^k\Big] = \frac{1}{2} - \frac{1}{2}\text{erf}\Bigg(\frac{g_j^k - \boldsymbol{h}_j^{kT}\bar{\boldsymbol{x}}_i^k}{\sqrt{2\boldsymbol{h}_j^{kT}\Sigma_{x_i^k}\boldsymbol{h}_j^k}}  \Bigg)
\end{align*}

After a RADMPC subproblem is solved, we record the collision probabilities between the subproblem vehicle and the temporal vehicle obstacles \emph{at the initial timestep}. We do not record the collision probabilities at other timesteps because RADMPC replans for the vehicle at the following timestep. 

After RADMPC converges on a solution, the maximum probability of collision of a vehicle with another temporal vehicle obstacle provides a direct metric for the probability of collision between the vehicle pair. This is used to determine vehicle pairs that have significant interactions and must be planned together. For every vehicle pair, there are two candidate maximum collision probabilities since both vehicles avoid the other's temporal vehicle obstacle when planning. The greater of the two probabilities is chosen as the maximum probability of collision for the vehicle pair. If the maximum probability of collision for the vehicle pair exceeds a given threshold $\Psi$, the vehicle pair must be planned in a coupled manner.

\subsection{Distributed Solving}
Determining coupled vehicle sets is analogous to determining the disconnected parts of a graph where vehicles are nodes and couplings are edges between vehicles. Path-planning problems are created for each coupled vehicle set. Applying IRA on every problem and combining the solutions computes the RADMPC solution to the original multi-vehicle problem.

\subsubsection{Computation Time Analysis}
We now analyze the solution time of our two-step approach. For ease of notation, we will factor out runtime due to common factors between our approach and the centralized approach (e.g. static obstacles used in both approaches). A centralized application of IRA to a problem with $N$ vehicles over $T$ timesteps uses $O(N^2T)$ binary variables for vehicle avoidance constraints between vehicle pairs, causing a $O(e^{N^2T})$ computational complexity.

The solution time of RADMPC likewise depends on the number of vehicles in the problem and the number of timesteps it takes for vehicles to reach their goal in our receding horizon approach. Since each decentralized subproblem uses temporal vehicle obstacles for vehicle avoidance, there are $O(NT)$ binary variables that imply $O(e^{NT})$ computation complexity per subproblem. Using the conservative assumption that RADMPC requires all $T$ timesteps to execute, there are $N$ decentralized problems over $T$ timesteps giving RADMPC $O(NTe^{NT})$ total computational complexity. 

Let us assume that analysis of the RADMPC approximation separates the full vehicle set into $\frac{N}{P}$ coupled vehicle sets with $P$ vehicles each. Applying centralized IRA on problems covering the coupled vehicle sets has $O(\frac{N}{P}e^{P^2T})$ computational complexity. Factoring in RADMPC runtime indicates $O(NTe^{NT} + \frac{N}{P}e^{P^2T})$ complexity. In the optimal scenario, we would be able to completely separate the problem into $N$ subproblems that are each individually solved with IRA for $O(NTe^{NT} + Ne^T)$ complexity. In the worst case, the complexity is $O(NTe^{NT} + e^{N^2T})$. However empirical results indicate that RADMPC decouples most vehicle couplings and generally does not exhibit worst case performance.

\section{Results}
In this section, we demonstrate empirical results of the RADMPC approach. We focus on analyzing the accuracy of the RADMPC approximation, average runtime of our approach compared to centralized IRA, and correctness of the RADMPC solution.

\subsection{RADMPC Approximations}
\begin{figure}[!htb]
\minipage{0.23\textwidth}%
\includegraphics[width=\linewidth]{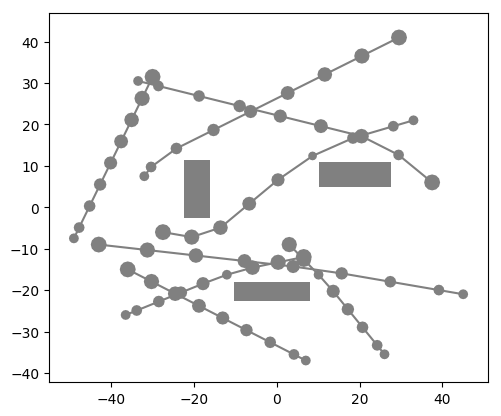}
\endminipage\hfill
\minipage{0.23\textwidth}
\includegraphics[width=\linewidth]{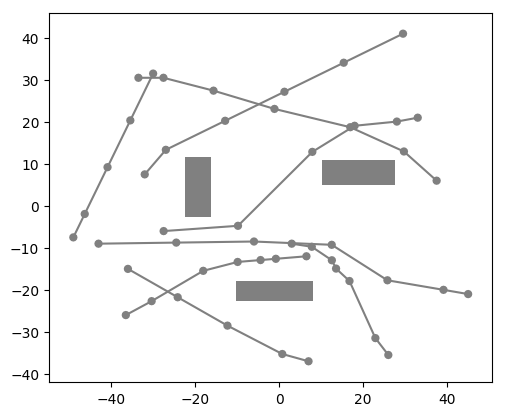}
\endminipage
\caption{Centralized solution on left, RADMPC approximation on right}
\label{solution_approximation}
\end{figure}

Figure \ref{solution_approximation} demonstrates the RADMPC approximation on an eight vehicle problem. The RADMPC approximation is a good approximation of the centralized solution with only slight disturbances A visual inspection of multiple sample problems indicates that RADMPC generally approximates the centralized solution well enough to be useful.

However, the accuracy of the RADMPC approximation decreases in some cases  especially when the complexity of the multi-vehicle problem increases. This may be because larger vehicle sets exhibit perturbances where "ripple" effects in RADMPC cause the displacement of a single vehicle to propagate to the other vehicles. Significant deviations from the centralized solution could lead RADMPC to make incorrect determinations of coupled vehicle sets, adversely affecting the correctness of the RADMPC solution. In general, RADMPC works the best in situations where vehicle interactions are less abundant i.e. situations where the problem should be decoupled.

\subsection{Runtime Comparisons}
We compare the average runtime of our approach and a centralized application of IRA by drawing 50 sample problems for each of $N \in [2, 8]$ vehicles and generating random vehicle starts and goals for each sample problem. We use a simple environment with three regular obstacle and use the following parameters:

\begin{align*}
    \boldsymbol{A} = 
    \begin{bmatrix} 
    1 & 0 & dt & 0\\ 
    0 & 1 & 0 & dt\\
    0 & 0 & 1 & 0\\
    0 & 0 & 0 & 1
    \end{bmatrix} &\quad
    \boldsymbol{B} = 
    \begin{bmatrix} 
    \frac{1}{2}dt^2 & 0\\ 
    0 & \frac{1}{2}dt^2\\
    dt & 0\\
    0 & dt
    \end{bmatrix}\\\\
    T = 10 \quad \Delta = 0.05 &\quad \Psi = 0.000001\\
    J(\boldsymbol{X},\boldsymbol{U}) &= \overset{N}{\underset{i = 0}{\sum}}\overset{T}{\underset{k = 0}{\sum}} |u_k^i|
\end{align*}

\begin{figure}[hb]
\begin{center}
\includegraphics[width=1.0\columnwidth]{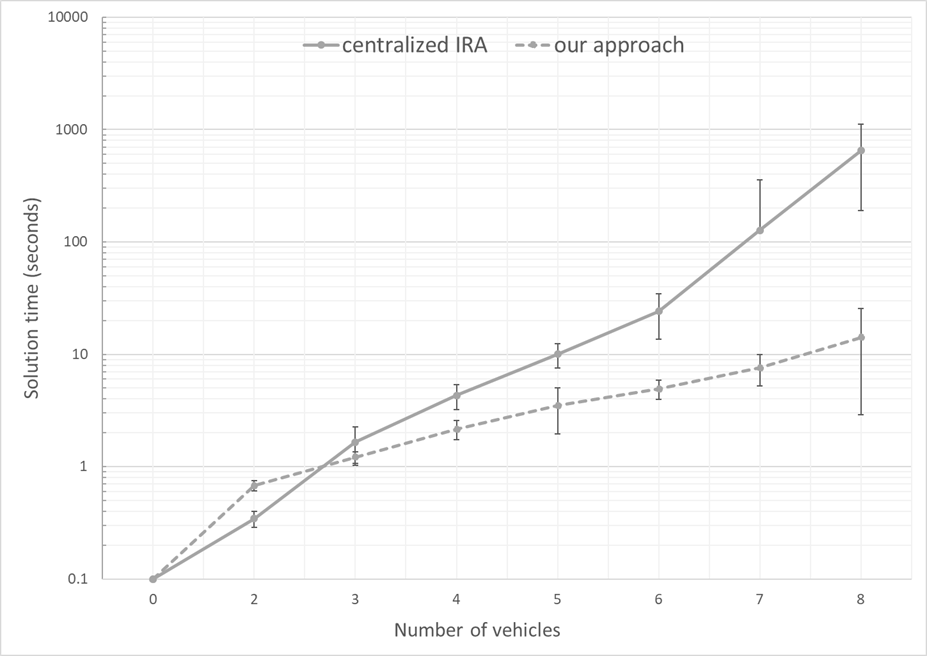}
\caption{Runtime comparison between centralized IRA and our approach}
\label{runtime_comparison}
\end{center}
\end{figure}

Figure \ref{runtime_comparison} demonstrates a significant reduction in solution time when solving complex RMPC problems with the heuristic, even when RADMPC runtime is included. As the size of the vehicle set increases, our approach will run faster by margins that increase with vehicle account. In our largest experiments involving 8 vehicles, the mean speedup factor reached 46. Much of this improvement can be attributed to the observation that most randomly generated complex problems can be separated into smaller, simpler problems with one or two vehicles each.

\subsection{Simulation}
A Monte Carlo simulation is used to verify the correctness of the RADMPC solution. Our Monte Carlo simulation uses the sample problems described in the prior section. A success rate is computed for each sample problem to express correctness. A correct solution is a solution that does not exhibit a probability of collision above the risk bound. To compute vehicle collisions in the RADMPC solution, every vehicle state $\boldsymbol{x}_k^i$ in the computed state sequence $\boldsymbol{X}$ is sampled 100,000 times. Vehicle collisions are detected using a simple computation that ensures that sampled vehicle states at corresponding time $k$ are not within two vehicle radii of each other.

\begin{figure}[h]
\begin{center}
\includegraphics[width=1.0\columnwidth]{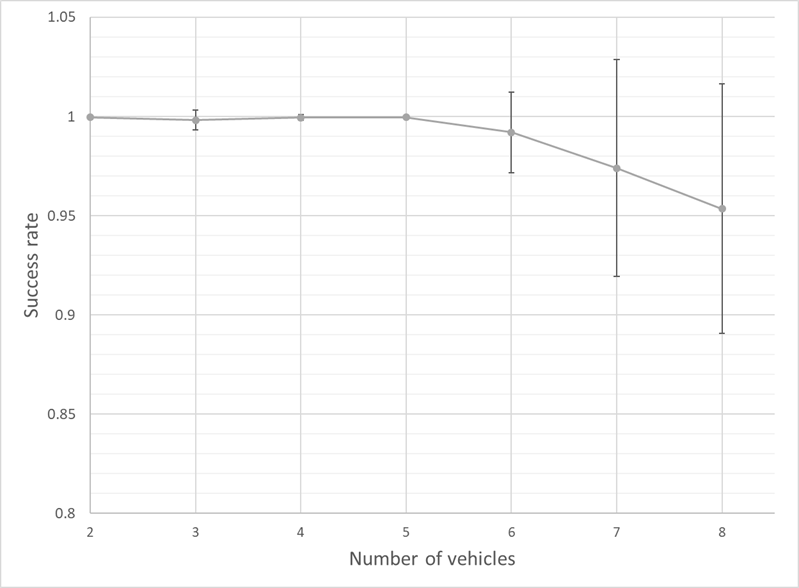}
\caption{Success rate of our approach with different numbers of vehicles}
\label{success_rates}
\end{center}
\end{figure}

Our approach on average works extremely well for the range of vehicles we considered for our experimentation. However, the success rate decreases as the number of vehicles increases and more deviation from the average success rate occurs. This trend can be expected to continue as the complexity of the problem increases, indicating that RADMPC is not as well suited for approximating complex problems with extensive inter-vehicle interaction. Still, the results presented demonstrate that our approach is useful to problems that include up to eight vehicles, with possible extension to more vehicles with further development.

\section{Conclusions}
We have presented a method that intelligently decouples computationally expensive chance-constrained multi-vehicle  problems to reduce solution time complexity. We describe a novel path-planning method called Risk-Aware Decentralized Model Predictive Control (RADMPC) to determine sets of coupled vehicles that have should be planned together. Applying IRA to each set generates collision-free paths in far less time than a centralized application of IRA.

We envision our approach being used to enable more rapid planning capabilities for vehicle swarms being used in the field. Although this approach was specifically designed to support autonomous underwater vehicle exploration, fast multi-vehicle path-planning spans a large range of academic, industrial, and humanitarian applications. Further research into fast multi-vehicle path-planning will facilitate more rapid integration of large-scale autonomous vehicle solutions in situations where they are sorely needed.

\section{Acknowledgement}
I would like to sincerely thank Benjamin Ayton for advising me on this project over the past year and helping me develop my skills as an undergraduate researcher. Thanks to Professor Brian Williams for being my faculty adviser and the MERS group for welcoming me into the family. Lastly, thanks to the MIT SuperUROP program for financially and logistically supporting this research project.

\end{document}